\documentstyle[prl,aps,floats,epsf]{revtex} 
\tighten 
 
\begin{document} 
\draft 
 
\title{ 
Semiclassical Theory of Coulomb Blockade Peak Heights in \\
Chaotic Quantum Dots 
} 
 
\author{Evgenii E. Narimanov,$^{1}$ Harold U. Baranger,$^{2}$ 
Nicholas R. Cerruti,$^{3}$  Steven Tomsovic$^{3}$ } 
 
\address{$^{1}$ Electrical Engineering Department, Princeton University,
Princeton NJ 08544}
\address{$^{2}$ Department of Physics, Duke University, Box 90305, 
Durham NC 27708-0305} 
\address{$^{3}$ 
Department of Physics, Washington State University, 
Pullman WA 99164-2814 
} 
 
\date{ \today} 
 
\maketitle

\begin{abstract} 
We develop a semiclassical theory of Coulomb blockade peak heights in 
chaotic quantum dots. Using Berry's conjecture, we calculate the  
peak height distributions and the correlation functions. We demonstrate 
that the corrections to the corresponding results of the standard  
statistical theory are non-universal and can be expressed in terms of  
the classical periodic orbits of the dot that are well coupled to the  
leads. The main effect is an oscillatory dependence of the peak heights
on any parameter which is varied; it is substantial for both symmetric 
and asymmetric lead placement. Surprisingly, these dynamical effects
do not influence the full distribution of peak heights, but are clearly
seen in the correlation function or power spectrum. For non-zero temperature,
the correlation function obtained theoretically is in good agreement
with that measured experimentally.
\end{abstract} 
\vspace*{-0.05 truein} 
\vspace*{-0.15 truein} 
 
\section{Introduction} 
 
The Coulomb blockade is a fundamentally classical effect in 
microstructures---the addition of an electron to an isolated microstructure 
requires a certain amount of electrostatic energy, the charging energy 
$e^2/2C$ where $C$ is the capacitance of the structure.  It is the simplest 
effect of electron charge in microstructures and has been extensively 
studied with regard to both fundamentals and applications in single electron 
transistors~\cite{nato-book}.  One common way to study the Coulomb blockade 
is by measuring the conductance through a nearly isolated nanoparticle 
(using tunneling contacts) as a function of a gate voltage which tunes the 
electrostatic potential of the particle. For most values of the gate 
voltage, the conductance is very small since the flow of electrons is 
blocked because the charging energy is not available. However, when the gate 
voltage is tuned so that states differing by one charge have the same energy, 
there is a peak in the conductance. The height of this peak is simply the 
conductance of the two tunnel barriers in series, and the spacing of the 
peaks is uniform with separation $e^2/C$. 
 
For the smallest quantum dots and at low temperature, however, quantum 
mechanical interference becomes important. Interference causes variation in 
both the height and spacing of the conductance peaks. For the spacing, 
single particle quantization and the residual interactions among the 
electrons are important. For the height, the nature of the wave functions 
become critical: if the wave function of the state at the chemical potential 
is poorly coupled to the leads---if it has nodes at the leads---then the 
conductance peak is small, but if the wave function is well coupled to the 
leads then the peak is large. In this paper, we restrict our attention to 
fluctuations in the conductance peak heights and investigate what this tells 
us about wave functions in quantum dots. 
 
Since dots are generally irregular in shape, the classical dynamics of the 
electrons is chaotic, and so the characteristics of Coulomb blockade peaks 
reflect those of wave functions in chaotic 
systems~\cite{jalabert,nato-book-2,Stopa98}.  Previously, a statistical 
theory for the peaks was developed\cite{jalabert,nato-book-2} by assuming 
these wave functions to be completely random and uncorrelated with each 
other.  The random matrix theory used was known to be a good description of 
energy level statistics, and so likely to be reasonable for wave functions. 
The experimental data~\cite{Chang96,MarcusFolk96} for the distribution of the 
Coulomb blockade peak heights were found to be in excellent agreement with 
the predictions of the statistical theory, thus supporting the conjecture of 
effective ``randomness'' of the quantum dot wave functions. 
 
A potential problem with the statistical theory was, however, evident in one
of the first experiments:  there is no correlation between 
different wave functions in
random matrix theory so the statistical theory predicts zero correlation
between neighboring conductance peaks, but in one of the
experiments~\cite{MarcusFolk96} correlation was clearly present in the form
of a slowly varying envelope modulating the peak heights.  In subsequent
years a number of different effects were investigated as candidates to
explain this correlation.  The simplest is the effect of nonzero temperature:
since excitation above the Fermi level is possible, several resonances
contribute to each peak and a given resonance contributes to several
neighboring conductance peaks, inducing correlation. However, in a detailed
study, this was found to be insufficient to account for the observed
correlations~\cite{Alhassid98}. Other explanations that were explored include
correlation due to spin-paired levels~\cite{Alhassid98,MarcusPatel98}, due to
a decrease of the effective level spacing found in density functional
calculations~\cite{Stopa96}, and due to level anticrossings in interacting
many-particle systems~\cite{Hackenbroich97}. While these latter explanations
rely on subtle electron-electron interaction effects, here we argue that peak
height correlations already arise within an effective single-particle picture
of the electrons in the quantum dot. The specific internal dynamics of the
dot, even though it is chaotic, modulates the peaks: because all systems have
short-time dynamical features, chaos is not equivalent to randomness.

While the statistical theory is ``universal'' in that it depends on no
specific features of the quantum dot at hand, the classical dynamics in the
dot is clearly not universal. Thus, while correlations between the
conductance peak heights are generally present in quantum dots, the
particular correlations in a given dot are not universal but rather involve
detailed information about the dot. The simplest information to include is
the spatial correlation function of the wave functions---this is very short
length dynamical information---and an approach including this effect was
given in Ref. \onlinecite{Vallejos}. Going beyond this, we use semiclassical
techniques to derive a relation between the quantum conductance peak height
and the classical periodic orbits in the dot.

The main result is that as a system parameter varies---the magnetic field,
for instance, or the number of electrons in the dot (controlled by varying a
gate voltage)---the interference around each periodic orbit oscillates
between being destructive and constructive. When the interference is
constructive for those periodic orbits which come close to the leads used to
contact the dot, the wave function is enhanced near the leads, the dot-lead
coupling is stronger, and so the conductance is larger. Likewise,
destructive interference produces a smaller conductance. The resulting
modulation at frequencies corresponding to the periodic orbits can be
substantial. Because of dephasing effects, only the short periodic orbits,
indeed perhaps only the shortest one, is likely to be significant.

Similar short-time dynamical effects have been noted in other contexts such
as atomic and molecular spectra~\cite{Gutzwiller,BeimsDelos98,heller},
eigenfunction scarring~\cite{heller,KaplanHeller}, magnetotransport in
antidot lattices~\cite{antidots}, and tunneling into quantum
wells~\cite{e2n_prl,monteiro,saraga98,e2n_physicad}. The periodic orbit
modulation that we discuss here is completely omitted in theories in which
the wave function is assumed to change randomly as the system
changes~\cite{jalabert,nato-book-2}. Reassuringly, the predicted dynamical
modulation is of the type in the original anomalous
experiment~\cite{MarcusFolk96}. More recently, other experimental data has
been published which show the effect~\cite{MarcusPatel98,MarcusCronenwett97},
but to date no systematic experimental study of this effect has been
performed.

In the rest of this paper, we generalize some results that have been
previously reported in Ref.~\onlinecite{Nar_cbdyn1} to address asymmetric
lead placement and to incorporate temperature dependence.  The derivation
given here is completely different from the previous one which relied on the
methods of Ref.~\onlinecite{e2n_physicad}: here our approach in terms of a
statistical ansatz for the wave functions yields more results for chaotic
systems but misses the results for regular systems that we obtained
previously.  It has been suggested that asymmetric lead placement would not
produce an observable oscillation in the average conductance~\cite{Vallejos},
but the method employed there only included spatial correlations in the wave
functions and not the short-time dynamics which we consider here.  In the
first section we express the height of the conductance peak in terms of the
resonant wave function. The basic ansatz for the distribution of the wave
functions, including dynamical effects, is presented in
Section~\ref{wavefuncts}. In Section~\ref{semiclG} our results for the
conductance peak heights are obtained. Comparison to numerical results for
the stadium billiard in Section~\ref{numerics} confirms the adequacy of the
semiclassical approach. Finally, we close with a summary and discussion of
future directions.

\section{The Height of a Conductance Peak in Coulomb Blockade}
\label{conductance}

Our starting point is the connection between the Coulomb blockade peak
heights and the widths of the levels in the quantum dot. This connection is
well-known~\cite{beenakker}; it allows us to express the conductance in terms
of single-particle quantities.  We consider a dot close to two leads so that
the width, $\Gamma$, of a level comes from tunneling of the electron to
either lead. When the mean separation of levels is larger than the
temperature $T$ which itself is much larger than the mean width, the
electrons pass through a single quantized level in the dot, and the
conductance peak height is~\cite{beenakker} 
\begin{eqnarray} 
G_{\rm peak} & =
& \frac{e^2}{h} \frac{\pi }{2 kT} 
\frac{\Gamma_1 \Gamma_2}{\Gamma_1 + \Gamma_2} \label{eq:g_max} 
\end{eqnarray} 
where $\Gamma_1$ and $\Gamma_2$ are
the partial decay widths due to the tunneling into a single lead, and spin
degrees of freedom are neglected.  In particular, when the leads are
identical and symmetrically attached to the dot, \begin{eqnarray} G_{\rm
peak} & = & \frac{e^2}{h} \frac{\pi }{4 kT} \Gamma_1 \label{eq:g_max_sym}
\end{eqnarray}

The partial width is related by Fermi's Golden Rule to the square of the
matrix element for tunneling between the lead and the dot,
$M^{\ell\rightarrow d}$. A convenient expression for the matrix element in
terms of the lead and dot wave functions, $\Psi_\ell$ and $\Psi_d$,
respectively, was derived by Bardeen~\cite{bardeen} and can be expressed
as~\cite{e2n_prl,monteiro} \begin{eqnarray}
   M^{\ell \rightarrow d} = \frac{\hbar^2}{m_*}  \int_{S} d{\bf r}
   \Psi_{\ell}({\bf r}) {\bf \bigtriangledown} \Psi_{d}({ \bf r})
   \label{eq:me_approx} \end{eqnarray} where the surface $S$ is the edge of
the quantum dot. The partial width, $\Gamma$, then, depends on the square of
the normal derivative of the dot wave function at the edge weighted by the
lead wave function.  The dot wave function $\Psi_d$ in (\ref{eq:me_approx})
is calculated for the effective potential, which accounts for interactions in
the dot in the mean-field approximation. For the partial width we then obtain
\begin{eqnarray}
   \Gamma_\alpha\left[\Psi_d\right] & = & \frac{2 \pi \hbar^4}{m_*^2}
   \sum_{\ell} \rho_\ell^{(\alpha)} \int_{S} d{\bf r_1} \cdot {\bf
   \bigtriangledown} \Psi_{d}\left({\bf r_1}\right) \int_{S} d{\bf r_2} \cdot
   {\bf \bigtriangledown} \Psi_{d}\left({\bf r_2}\right)^* \left[
   \Psi_{\ell}^{(\alpha)}\left({\bf r_1}\right)^*
   \Psi_{\ell}^{(\alpha)}\left({\bf r_2}\right) \right]
   \label{eq:gamma_partial} \end{eqnarray} where $\alpha$ is the index of the
lead, the integer $\ell$ represents different transverse subbands in the
lead, and $\rho_\ell$ is the density of states in the lead for a given
subband. To obtain the statistics of the conductance peak heights, we thus
need to know the statistical properties of the dot wave functions $\Psi_d$.

\section{Wave Functions in the Dot: The Statistical Description}
\label{wavefuncts}

For a {\it single} dot, we consider an ensemble of Coulomb blockade
peaks---measured either in a narrow interval of gate voltage or obtained by
following a single resonance under continuously changing magnetic field. The
wave functions associated with the peaks of the conductance will vary---or
``fluctuate''---in a way characterized by a distribution $P[\psi]$ which we
seek.

It was first conjectured by M.~V.~Berry that the wave functions of a
classically chaotic system fluctuate with certain universal properties and
can be characterized as random variables~\cite{Berry77}. This is the
foundation of the first statistical theory of peak heights \cite{jalabert}.
Subsequently, the statistical ansatz made by Berry has been further
developed. One direction of refinement is the incorporation of some short
length-scale aspects of the real classical dynamics. First, a constraint of
an arbitrary correlation function \begin{equation}
   C\left({\bf r}_1,{\bf r}_2\right) \equiv \int {\cal D}\psi
   P\left[\psi\right] \psi^*\left({\bf r}_1\right) \psi\left({\bf r}_2\right)
   \label{eq:corr} \end{equation} was incorporated into the
ansatz~\cite{Alhassid95,Srednicki96}. By using the correlation function of a
random superposition of plane waves, the probability distributions of
level-widths and conductance peaks in the case of multi-mode leads to the
quantum dot were found~\cite{Vallejos,Alhassid95}. 
A distribution similar to this
ansatz was derived microscopically for disordered systems, a specific kind of
chaotic system, using the nonlinear sigma
model~\cite{Prigodin93,Prigodin94,Prigodin95,PrigodinSridhar95}.

The next step was to constrain the correlation function by the short-time
classical dynamics. Using the short-path semiclassical correlation, Srednicki
and coworkers~\cite{SrednickiStiernelof96,Hortikar98} studied correlations in
chaotic eigenfunctions at large separations and found that the predicted
correlations are in excellent agreement with numerical calculations in
chaotic billiards~\cite{SrednickiStiernelof96}. This semiclassically
constrained ansatz for $P[\psi]$ is much harder to justify---certainly no
derivation in disordered systems can be made. However, progress towards this
goal has been achieved by Kaplan and Heller by treating the nonlinear effects
of classical recurrences~\cite{KaplanHeller}. In a recent paper by
Kaplan~\cite{Kaplan} short-time dynamics were incorporated into the general
probability distribution of Ref.~\onlinecite{Alhassid95} to improve the
random matrix theory results for the conductance peak height statistics.

Here we use a maximum entropy technique~\cite{maxent} to derive the specific
form of the distribution $P[\psi]$ that we need. An advantage of this
approach is that arbitrary constraints can be introduced, as in the case of
normalization which we discuss below.  We make the following ansatz:  the
distribution $P\left[\psi\right]$ maximizes the information
entropy~\cite{Shannon48} \begin{equation}
   H = - \int {\cal D}\psi \ P\left[\psi\right] \log P\left[\psi\right]
   \label{eq:entropy} \end{equation} within the space allowed by the
constraints.  Here the measure corresponding to the distribution
$P\left[\psi\right]$ is defined in the standard way~\cite{Hortikar98}
\begin{equation}
   {\cal D}\psi_d = \lim_{N \to \infty} \Pi_{n = 1}^N d\psi_d\left( {\bf r}_n
   \right) \label{eq:Dpsi} \end{equation} so that the product
$P\left[\psi_d\right] {\cal D}\psi_d$ represents the probability that a wave
function $\psi\left({\bf r}\right)$ of the original ensemble is between
$\psi_d\left( {\bf r} \right)$ and $\psi_d\left( {\bf r} \right) +
d\psi_d\left( {\bf r} \right)$ for any point ${\bf r}$ inside the dot.

Assuming that the {\it only} constraint imposed on the ensemble of wave
functions is the correlation function $C\left({\bf r}_1,{\bf r}_2\right)$,
the maximum of the functional (\ref{eq:entropy}) under the constraint
(\ref{eq:corr}) is equivalent to the extremum of the functional
\begin{eqnarray}
   F\left[\psi\right] & = & \int {\cal D}\psi \ \left[ -P\left[\psi\right]
   \log P\left[\psi\right] - \int d{\bf r}_1 \int d{\bf r}_2 \
   \lambda\left({\bf r}_1,{\bf r}_2\right) \bigl\{ \psi^*\left({\bf
   r}_1\right) \psi\left({\bf r}_2\right) P\left[\psi\right] - C\left({\bf
   r}_1,{\bf r}_2\right) \bigr\} \right] \label{eq:functional} \end{eqnarray}
where the Lagrange multiplier $\lambda\left({\bf r}_1,{\bf r}_2\right)$ can
then be determined from Eq. (\ref{eq:corr}). Setting the first variation of
$F\left[\psi\right]$ equal to zero, we find that $P[\psi]$ is Gaussian. The
final result, obtained by substituting Eq. (\ref{eq:corr}) to find
$\lambda\left({\bf r}_1,{\bf r}_2\right)$, is \begin{eqnarray}
   P\left[\psi\right] & = & A \exp\left[ - \frac{\beta}{2} \int d{\bf r}_1
   \int d{\bf r}_2 \ \psi^*\left({\bf r}_1\right) C^{-1}\left({\bf r}_1,{\bf
   r}_2\right) \psi\left({\bf r}_2\right) \right] \label{eq:p_psi_lambda}
\end{eqnarray} where $A$ is the normalization [independent of $\psi\left({\bf
r}\right)$], and $C^{-1}$ is the functional inverse of the two-point
correlation function $C\left( {\bf r}_2, {\bf r}_1 \left|
\varepsilon\right)\right.$ \begin{eqnarray}
   \int d{\bf r}_3 \ C^{-1}\left({\bf r}_1,{\bf r}_3\right) C\left({\bf
   r}_3,{\bf r}_2\right) & = & \delta\left({\bf r}_1 - {\bf r}_2\right) \;.
   \label{eq:func_inverse} \end{eqnarray} The coefficient $\beta \!=\! 1$ for
a system with time-reversal invariance, when the wave functions can be chosen
real, and  $\beta \!=\! 2$ otherwise.

It has been shown~\cite{Hortikar98} that in the small-$\hbar$ limit for
classically chaotic systems, the correlation function $C\left( {\bf r}_2,
{\bf r}_1 \right)$ can be expressed in terms of the semiclassical
approximation to the Green function\cite{Gutzwiller} $G_{\rm sc}\left( {\bf
r}_2, {\bf r}_1 \right)$ as \begin{eqnarray}
   C\left( {\bf r}_2, {\bf r}_1 \right) & = & \frac{1}{\pi
   \overline{\rho}_{\rm sc}} {\rm Im} G_{\rm sc}\left( {\bf r}_2, {\bf r}_1
   \right) + {\cal O}\left(\hbar^{3/2}\right) \label{eq:Csc} \end{eqnarray}
where $\overline{\rho}_{\rm sc}\left(\varepsilon\right)$ is the smooth part
of the density of states (DOS) in the dot, given by the leading order
(Thomas-Fermi) semiclassical approximation to the DOS.

In the semiclassical approximation, the energy-averaged Green function can be
expressed in terms of the classical trajectories (labeled by the index
$j$)~\cite{Gutzwiller,Reichl}, \begin{eqnarray}
   G_{\rm sc}\left( {\bf r}_2, {\bf r}_1 \right) & = & G_0\left({\bf r}_2,
   {\bf r}_1\right) + \frac{1}{i\hbar} \frac{1}{\sqrt{2 \pi i \hbar}}
    \sum_{j} \sqrt{\left| D_j \right|} \exp\left(i \frac{S_j}{\hbar} - i n_j
    \frac{\pi}{4} \right) \exp\left(-\frac{\tau^2_j W^2}{2 \hbar^2} \right)
    \;, \label{eq:Gsc} \end{eqnarray} where $S_j = S_j\left( {\bf r}_2, {\bf
r}_1 \right)$ is the classical action, $\tau_j$ is the period, the integer
$n_j$ is the topological index~\cite{Gutzwiller} of the trajectory $j$, and
the amplitude $ D_j$ is \begin{eqnarray}
    D_j & = & {\rm det}\left( \begin{array}{cc} 
      \frac{\partial^2 S_j\left( {\bf r}_2, {\bf r}_1 \right)}
	   {\partial {\bf r}_2 \ \partial {\bf r}_1} & \frac{\partial^2
      S_j\left( {\bf r}_2, {\bf r}_1 \right)}
	   {\partial \varepsilon \ \partial {\bf r}_1} \\ \frac{\partial^2
      S_j\left( {\bf r}_2, {\bf r}_1 \right)}
	   {\partial \varepsilon \ \partial {\bf r}_2} & \frac{\partial^2
      S_j\left( {\bf r}_2, {\bf r}_1 \right)}
	   {\partial \varepsilon^2 } \end{array} \right) \;.  \label{eq:D}
\end{eqnarray} We have specialized to two spatial dimensions and the last
exponential in Eq.~(\ref{eq:Gsc}) is due to a Gaussian averaging over an
energy window of width $W$ described below.  The function $G_0\left( {\bf
r}_2, {\bf r}_1 \right)$ is the contribution of the non-classical so-called
``zero-length'' trajectories, those with actions less than or of order
$\hbar$. Therefore, $G_0\left({\bf r}_2, {\bf r}_1\right)$ cannot be obtained
using the stationary-phase approximation, but may be
evaluated~\cite{Berry77,Hortikar98} by replacing the actual propagator
$\langle {\bf r}_2 \left| \exp(- i H t / \hbar ) \right|  {\bf r}_1 \rangle$
by its free space analog \begin{eqnarray}
   \langle {\bf r}_2 \left| \exp\left(- \frac{i H t}{\hbar} \right) \right|
   {\bf r}_1 \rangle \approx \int \frac{d  {\bf p} }{\left(2 \pi
   \hbar\right)^2} \exp\left(i \frac{ {\bf p}\left(  {\bf r}_2 - {\bf r}_1
   \right)}{\hbar} \right) \exp\left(- \frac{i H\left( {\bf p},{\bf r}_0
   \right) t}{\hbar} \right) \end{eqnarray} where ${\bf r}_0 \equiv \left(
{\bf r}_2 + {\bf r}_1 \right) / 2$.  The corresponding Green function is then
\begin{eqnarray}
   G_0\left({\bf r}_2, {\bf r}_1 \right)  & = & \int \frac{d  {\bf p}
   }{\left(2 \pi \hbar\right)^2} \exp\left(i \frac{ {\bf p} \cdot
	\left( {\bf r}_2 - {\bf r}_1 \right)}{\hbar} \right)
   \frac{1}{\varepsilon - H\left({\bf p},{\bf r}_0 \right) + i 0} \;.
\end{eqnarray} Note that because of the short trajectory involved, this part
of the Green function varies very smoothly as a function of energy.  The
smooth part of the correlation function which results is \begin{eqnarray}
   C_0\left({\bf r}_2, {\bf r}_1 \right)  & = & \frac{1}{\overline{\rho}_{\rm
   sc}} \int \frac{d  {\bf p} }{\left(2 \pi \hbar\right)^2} \cos\left(\frac{
   {\bf p}\left(  {\bf r}_2 - {\bf r}_1 \right)}{\hbar} \right) \delta\bigl(
   \varepsilon - H\left({\bf p},{\bf r}_0 \right)  \bigr) \;, \end{eqnarray}
and so $C_0\left({\bf r}_2, {\bf r}_1 \right) \propto J_0\left( p \left| {\bf
r}_2 - {\bf r}_1 \right| / \hbar  \right)$.  This smooth part of the
correlation function is rather local in that it decays monotonically with
separation.  Thus, having fully specified the correlation function we wish to
use, we finally obtain \begin{eqnarray}
P\left(\psi_d\left|\varepsilon\right)\right. & \sim & \exp\left[ -
\frac{\beta}{2} \int d{\bf r}_1  \int d{\bf r}_2 \psi^*\left( {\bf r}_1
\right) G_{\rm sc}^{-1}\left({\bf r}_2, {\bf r}_1 \left|
\varepsilon\right)\right.  \psi\left( {\bf r}_2 \right) \right] \;.
\label{eq:P_psi} \end{eqnarray}

A few remarks are required about the width of the energy window $W$.  
In the semiclassical limit there arises an increasingly broad separation
between the short-time dynamics that give rise to system specific behavior
and the long orbits that are responsible for generating universal statistical
fluctuations~\cite{KaplanHeller}.  The width $W$ is 
chosen such that the short periodic orbits are included in the sum
essentially undamped whereas the long orbits are eliminated since
their contributions are already accounted for in the statistical ansatz.  For 
the rest of this paper, we will eliminate the explicit dependence on $W$ 
and the sum is understood to contain only the linear dynamics.      
 
The general ensemble defined by the distribution (\ref{eq:p_psi_lambda}) has, 
however, certain limitations. Strictly speaking, in its general form this 
ensemble is only suitable for calculations of those observables which can be 
represented in terms of only two-point products $\psi^*\left({\bf r}_1\right) 
\psi\left({\bf r}_2\right)$. The reason for this problem is as follows: 
instead of the proper normalization of {\it each member of the ensemble}, 
\begin{eqnarray} 
   \int d{\bf r} \ \left| \psi\left({\bf r}\right) \right|^2 & = & 1 \;, 
\end{eqnarray} 
the normalization of the wave functions is satisfied {\it only on average}, 
\begin{eqnarray} 
   \int {\cal D}\psi\left({\bf r}\right) \ 
   P\left[\psi\left({\bf r}\right)\right]  
   \int d{\bf r} \ \left| \psi\left({\bf r}\right) \right|^2 & = & 1 \;. 
\end{eqnarray} 
As a result, the higher order moments,  
$\Delta_n \!\equiv\! \langle \int d{\bf r}_1 \ldots \int d{\bf r}_n  
\left|\psi\left({\bf r}_1\right)\right|^{2} \ldots 
\left|\psi\left({\bf r}_n\right)\right|^{2} \rangle_\psi$, 
of the distribution are different from unity. Therefore, in its general from, 
the ensemble defined by (\ref{eq:p_psi_lambda}) is not suitable for 
calculations which are sensitive to the $n \!>\! 1$ moments of the distribution 
$P\left[\psi\right]$, such as for the description of the residual 
interactions in quantum dots~\cite{BroOreHal,BarUllGla,UllBar_inprep}. 
 
The method developed in this section yields a straightforward 
way to generalize the distribution (\ref{eq:p_psi_lambda}) to properly 
account for the higher moments. For example, adding an additional constraint 
\begin{eqnarray} 
   \int {\cal D}\psi P\left[\psi\right] 
   \int d{\bf r}_1 \int d{\bf r}_2 \  
   \left| \psi\left({\bf r}_1\right) \right|^2 
   \left| \psi\left({\bf r}_2\right) \right|^2 
    & = & 1, 
   \label{eq:moment_2} 
\end{eqnarray} 
to the variational problem (\ref{eq:entropy}) will yield a generalization of 
the distribution (\ref{eq:P_psi}) which properly accounts for the moment 
$\Delta_2$. 
 
Note in contrast that the errors in the higher moments, $n \!>\! 1$, produced 
by the {\it semiclassical} distribution (\ref{eq:P_psi}) are of higher order 
in $\hbar$, $\delta_n \sim {\cal O}\left(\hbar^2\right)$, than the terms 
taken into account in $G_{\rm sc}$~\cite{Srednicki_pr_com}. As long as these 
higher-order corrections are not relevant for the quantity under 
consideration, one can generally use the {\it semiclassical} distribution 
(\ref{eq:P_psi}).

\section{Peak Heights Distribution} 
\label{semiclG} 
 
Since the Coulomb blockade peak heights are uniquely determined by the 
corresponding dot wave functions $\psi_d$, the peak heights distribution 
function $P\left(G\right)$ is given by 
\begin{eqnarray} 
   P\left(G\right) & = &  
   \int {\cal D}\psi_d \ P\left(\psi_d\right) 
   \delta\left(G - G_{\rm peak}\left[\psi_d\right]\right) 
   \label{eq:Pg} 
\end{eqnarray} 
where $G_{\rm peak}\left[\psi\right]$ is determined by Eqs. 
(\ref{eq:g_max}), (\ref{eq:g_max_sym}), and (\ref{eq:gamma_partial}). 
The width $\Gamma$ depends only on the wave function near the boundary of the 
quantum dot, as follows from Eq. (\ref{eq:gamma_partial}).  If the function 
$P_S\left(\bar{\psi}\right)$ represents the distribution of the wave functions 
in a narrow strip $S$ along the boundary of the quantum dot, so that 
\begin{eqnarray} 
   \psi\left({\bf r}\right) = \left\{ 
   \begin{array}{cc} 
      \bar{\psi}\left({\bf r}\right), & {\bf r} \in S \\ 
      \hat{\psi}\left({\bf r}\right), & {\bf r} \notin S  
   \end{array} 
   \right. \;, 
   \label{eq:psi_S} 
\end{eqnarray} 
then the conductance distribution is 
\begin{eqnarray} 
   P\left(G\right) & = &  
   \int {\cal D}\bar{\psi} \ P_S\left[\bar{\psi}|\varepsilon\right] 
   \delta\left(G - G_{\rm peak}\left[\bar{\psi}\right]\right) \;. 
   \label{eq:Pg_S} 
\end{eqnarray} 
The ``edge'' distribution $P_S$ can be obtained from the general distribution 
$P\left[\psi\right]$ by integrating out the values of $\hat{\psi}$, 
\begin{eqnarray} 
   P_S\left[\bar{\psi}\right] & = & \int {\cal D}\hat{\psi} \ 
   P\left[\psi\left\{\bar{\psi}, \hat{\psi}\right\}\right] \;. 
\end{eqnarray} 
As the distribution $P\left[\psi\right]$ is Gaussian, the resulting functional 
integral can be calculated exactly, yielding 
\begin{eqnarray} 
   P_S\left[\bar{\psi}\right] & = & A_S 
   \exp\left[ - \frac{\beta}{2} \int_S d{\bf q}_1 \int_S d{\bf q}_2   
   \bar{\psi}^*\left({\bf q}_1\right) 
   \bar{K}\left({\bf q}_1,{\bf q}_2\right)  
   \bar{\psi}\left({\bf q}_2\right)  
   \right]  
\end{eqnarray} 
where  
\begin{eqnarray} 
   \bar{K}\left({\bf q}_1,{\bf q}_2\right) & = & 
   C^{-1}\left({\bf q}_1,{\bf q}_2\right) 
   + 
   \int_{\Omega\backslash S} d{\bf q}_3   
   \int_{\Omega\backslash S}  d{\bf q}_4 
   C^{-1}\left({\bf q}_1,{\bf q}_3\right) 
   C\left({\bf q}_3,{\bf q}_4\right) 
   C^{-1}\left({\bf q}_4,{\bf q}_2\right) 
   \label{eq:K} 
\end{eqnarray}  
and $A_S$ is the new normalization constant. The spatial integrals are over 
the part of the total space $\Omega$ which is orthogonal to the edge $S$,
denoted $\Omega \backslash S$. 
 
As follows from Eqs. (\ref{eq:Csc})-(\ref{eq:Gsc}), the ``non-diagonal'' part 
of the correlation function is of a higher order in $\hbar$, $\sim {\cal 
O}(\sqrt{\hbar})$, compared to the ``diagonal'' part, $C_0 \sim  {\cal 
O}(1)$.  The second term in Eq. (\ref{eq:K}) involves the correlation 
functions $C\left({\bf q}_1,{\bf q}_3\right)$ and $C\left({\bf q}_4,{\bf 
q}_2\right)$, taken between the points of the {\it different} parts of the 
dot, the edge strip $S$ for one coordinate and the internal region 
$\Omega\backslash S$ for the other. It is therefore of higher order in 
$\hbar$, $\sim {\cal O}(\hbar)$, than the first contribution, $C^{-1}({\bf 
q}_1,{\bf q}_2)$ $\sim  \delta({\bf q}_2- {\bf q}_1){\cal O}(1) + {\cal 
O}(\sqrt{\hbar})$. Keeping such higher-order terms is not consistent with the 
leading-order semiclassical approximation we used for $C\left({\bf q}_1,{\bf 
q}_2\right)$. We therefore obtain 
\begin{eqnarray} 
   P_S\left[\bar{\psi}\right] & = & A_S 
   \exp\left[ - \frac{\beta}{2} \int_S d{\bf q}_1 \int_S d{\bf q}_2 \  
   \bar{\psi}^*\left({\bf q}_1\right) 
   C^{-1}\left({\bf q}_1,{\bf q}_2\right)  
   \bar{\psi}\left({\bf q}_2\right)   
   \right] \;. 
   \label{eq:P_edge} 
\end{eqnarray} 
An alternative to the argument given here proceeds by noting that integrating 
out $\hat{\psi}$ should yield a Gaussian in $\bar{\psi}$, and that this 
Gaussian, by construction of the ensemble, must reproduce the correct two 
point correlation function $C\left({\bf q}_1,{\bf q}_2\right)$. This 
alternative argument~\cite{Srednicki96} immediately yields the functional 
form (\ref{eq:P_edge}). 
 
When the {\it closed} dot is defined by the Dirichlet boundary conditions, 
the wave function in the narrow strip  $S$ near the ``edge''  
can be represented as 
\begin{eqnarray} 
   \bar{\psi} & = & z \varphi\left(y\right) 
\end{eqnarray} 
where $y$ is the coordinate along the boundary of the dot and $z$ is in 
the direction of the normal. In this limit, the correlation function is 
\begin{eqnarray} 
   C\left({\bf q}_1,{\bf q}_2\right) & = & z_2 z_1 \partial_n  
   C\left({y}_1, {y}_2\right) 
\end{eqnarray} 
where $ \partial_n C\left({y}_1, {y}_2\right)$ is defined as the correlation 
function of the normal derivatives of the wave function at the boundary of the 
dot and can be obtained as 
\begin{eqnarray} 
   \partial_n C\left( {y}_2, {y}_1 \right) & = &  
   \frac{1}{\pi \overline{\rho}_{\rm sc}} {\rm Im} 
   \partial_n  
   G_{\rm sc}\left( {y}_2, {y}_1 \right) 
   + {\cal O}\left(\hbar^{\left(d+1\right)/2}\right) 
   \label{eq:dCsc} 
\end{eqnarray} 
where  
\begin{eqnarray} 
   \partial_n G\left( {y}_2, {y}_1 \left|\right. \varepsilon\right) & \equiv & 
   \sum_m 
   \frac{\partial_n\psi_m^*\left(y_2,0\right) \partial_n\psi_m\left(y_1,0\right)} 
   {\varepsilon_m - \varepsilon + i0} \;. 
   \label{eq:dG}  
\end{eqnarray} 
The semiclassical approximation $\partial_nG_{\rm sc}$ for the normal 
derivative Green function was derived in Ref. \onlinecite{e2n_physicad}: 
\begin{eqnarray} 
   \partial_nG_{\rm sc}\left( y_2, y_1 \right) 
   & = & \partial_nG_0\left(y_2, y_1\right) \nonumber \\ 
   & + & \frac{4}{i\hbar^3} \frac{1}{\sqrt{2 \pi i \hbar}} 
   \sum_{j}  
   \left[{\bf p}_j\left(y_1\right)\right]_n 
   \left[{\bf p}_j\left(y_2\right)\right]_n 
    \sqrt{\left| D_j \right|} 
    \sin\left(\frac{S_j}{\hbar} - \bar{n}_j \frac{\pi}{4} \right) 
   \label{eq:dGsc} 
\end{eqnarray} 
where $\bar{n}_j$ and $[{\bf p}_j]_n$ are, respectively, the Maslov 
indexes~\cite{Gutzwiller} and the normal component of the classical momentum 
of the trajectory $j$.
%
%
%
%
 
In order to connect the dot wave functions to the lead, let 
$\{\phi_m\left(y\right)\}$ be the complete orthogonal set of the 
wave functions corresponding to the transverse potential of the lead.  Using 
this basis, we represent the function $\varphi\left(y\right)$ as 
\begin{eqnarray} 
   \varphi\left(y\right) & = & \sum_{m=0}^{\infty}  
                               a_m \phi_m\left(y-y_\ell\right) 
\end{eqnarray} 
where $y_\ell$ is the contact point of the lead.  Assuming that the tunneling 
between the lead and the dot is dominated by the contribution of the lowest 
transverse subband of the lead and using (\ref{eq:gamma_partial}) for the 
partial width $\Gamma_\alpha$, we obtain 
\begin{eqnarray} 
   \Gamma_\alpha & = &  
   {2 \pi}\rho^{(\alpha)}_0 \left(\frac{\hbar^2}{m_*}\right)^2 
   \left| \int dy \ \phi_0\left(y\right) \sum_m a_m \phi_m\left(y\right)\right|^2 
   = \frac{2 \pi \hbar^4}{m_*^2} \rho^{(\alpha)}_0 \left|a_0\right|^2 
   \label{eq:gamma_a} 
\end{eqnarray} 
where $\rho^{(\alpha)}_0$ is the density of states in the lead corresponding 
to the lowest transverse subband.  For an arbitrary moment of the partial 
width $\langle \Gamma_\alpha^m \rangle$ we therefore find 
\begin{eqnarray} 
   \langle \Gamma_\alpha^m \rangle & \propto & 
   \int da_0 \left[ \frac{2 \pi \hbar^4}{m_*^2} \rho^{(\alpha)}_0\right]^m 
   \left| a_0 \right|^{2 m}  
   \exp\left[ 
   - \left| a_0 \right|^2 
   \int dy_1 \int dy_2 \ 
   \phi_0\left(y_1\right)  
   \frac{1}{\partial_n C\left(y_1, y_2\right)} 
   \phi_0\left(y_2\right)  
   \right] \;. 
\end{eqnarray} 
 
To give explicit expressions for the distribution of level widths and 
conductance, we specialize to the time-reversal symmetric case ($\beta \!=\! 
1$, GOE) for the rest of this paper; the case when time-reversal symmetry is 
broken by a magnetic field ($\beta \!=\! 2$, GUE) can be treated in an 
analogous way.  In the presence of time-reversal symmetry, the wave functions, 
and hence the coefficient $a_0$, can be chosen real, yielding 
\begin{eqnarray} 
   \langle \Gamma^m \rangle & \propto &  
   \int d\Gamma \ \Gamma^m  
   \frac{\exp\left( - \Gamma / 2 \overline{\Gamma} \right)} 
   {\sqrt{\Gamma}} 
   \label{eq:eq_PT} 
\end{eqnarray} 
where 
\begin{eqnarray} 
   \overline{\Gamma} & = & \frac{\hbar^4}{m_*^2} 
   \frac{\rho^{(\alpha)}_0}{\bar{\rho}_{\rm sc}}  
   { \left( \int dy_1 \int dy_2 \ 
   \phi_0\left(y_1\right)  
   \frac{1} 
   {{\rm Im}\left[\partial_n G_{\rm sc}\left(y_2, y_1\left|\varepsilon 
   \right.\right)\right]} 
   \phi_0\left(y_2\right) \right) }^{-1} \;. 
   \label{eq:gamma_local_average} 
\end{eqnarray} 
{\it Thus the partial width is characterized by the Porter-Thomas 
distribution } 
\begin{eqnarray} 
   P\left(\Gamma\right) \propto  
   \frac{1}{\sqrt{\Gamma}} 
   \exp\left( - \frac{\Gamma}{2\overline{\Gamma}}\right) 
\end{eqnarray} 
{\it with the slowly varying local average 
$\overline{\Gamma}\left(\varepsilon\right)$.} This explicit result for the 
distribution of level widths is the main result of this section. 
 
The conductance distribution $P\left(G\right)$ can now be simply derived in 
two limiting cases: (i) when the leads are placed symmetrically, so that 
$\Gamma_1 \!=\! \Gamma_2$ [cf. Eq. (\ref{eq:g_max_sym})], and (ii) when one 
of the partial widths is substantially smaller than the other, $\Gamma_1 
\!\ll\! \Gamma_2$. In both these cases $G \sim \Gamma_1$ [as follows from Eq. 
(\ref{eq:g_max})], and  the conductance distribution is also Porter-Thomas. 
The ``local average'' conductance, $\overline{G}$, is given by 
\begin{eqnarray} 
   \overline{G}\left(\varepsilon\right) & = &  
   \frac{e^2 \pi}{2 \gamma h kT} 
   \overline{\Gamma}\left(\varepsilon\right) 
\end{eqnarray} 
where the ``local average'' width $\overline{\Gamma}\left(\varepsilon\right)$ 
is defined by Eq. (\ref{eq:gamma_local_average}), and $\gamma \!=\! 1$ for 
$\Gamma_1 \!\ll\! \Gamma_2$, while $\gamma \!=\! 2$ for $\Gamma_1 \!=\! 
\Gamma_2$. 
 
In the general case, $\Gamma_1/\Gamma_2 \sim {\cal O}\left(1\right)$ but not 
identical, however, an exact calculation of the conductance distribution is 
complicated by the essentially nonlinear dependence of the conductance on the 
partial widths $\Gamma_1$ and $\Gamma_2$.  In order to calculate the actual 
conductance, we choose the area $S$ as the composition of two narrow strips, 
$S_1$ and $S_2$, near each of the leads. Using the transverse lead 
wave functions as the basis in each of the two strips, 
\begin{eqnarray} 
   \bar{\psi}\left(y,z\right) & = & z \left[  
   \sum_m a_m^{(1)} \phi^{(1)}_m\left(y-y_\ell^{(1)}\right) 
   +  \sum_m a_m^{(2)} \phi^{(2)}_m\left(y-y_\ell^{(2)}\right) 
   \right] 
\end{eqnarray} 
where the coordinates $y_\ell^{(1)}$ and $y_\ell^{(2)}$ represent the 
``contact points'' of the leads.  The partial widths $\Gamma_1$ and 
$\Gamma_2$ are then given by 
\begin{eqnarray} 
   \Gamma_1 & = & \frac{2 \pi \hbar^4}{m_*^2} \rho_0^{(1)} 
   \left| a_0^{(1)} \right|^2 \\ 
   \Gamma_2 & = & \frac{2 \pi \hbar^4}{m_*^2} \rho_0^{(2)} 
   \left| a_0^{(2)} \right|^2 \;. 
\end{eqnarray} 
Assuming equal density of states in the leads, $\rho_0^{(1)} (\varepsilon)  
\!=\! \rho_0^{(2)} (\varepsilon)$, for the conductance $G$ we obtain 
\begin{eqnarray} 
   G & = & \frac{\pi}{2}  
   \frac{e^2 \hbar^3 }{ m_*^2 k T }  \rho_{0} 
   \frac{|a_0^{(1)}|^2 |a_0^{(2)}|^2} 
        {|a_0^{(1)}|^2 + |a_0^{(2)}|^2} \;. 
\end{eqnarray}  
 
An arbitrary $n^{\rm th}$ moment of the conductance $G$, $\langle 
G^n\rangle$, can now be calculated by integrating over the coefficients  
$\{ a_m \}$ for $m\neq 0$, yielding 
\begin{eqnarray} 
   \langle G^n\rangle & \propto & 
   \int da_0^{(1)} \int da_0^{(2)} 
   \left[\frac{\left|a_0^{(1)}\right|^2 \left|a_0^{(2)}\right|^2} 
        {\left|a_0^{(1)}\right|^2 + \left|a_0^{(2)}\right|^2} 
   \right]^n 
   \exp\left[ 
   - {a_0^{(1)}} {\cal A}_{11} {a_0^{(1)}} 
   - {a_0^{(2)}} {\cal A}_{22} {a_0^{(2)}} 
   - 2  {a_0^{(1)}}^* {\cal A}_{12} {a_0^{(2)}}  
   \right] 
   \label{eq:g_distribution_A} 
\end{eqnarray} 
where the matrix ${\cal A}$ is 
\begin{eqnarray} 
   {\cal A}_{\alpha \beta} & = &  
   \int dy_1 \int dy_2 \  
   \phi_0\left(y_1 - y_{\ell}^{(\alpha)}\right) 
   \partial_n C^{-1}\left(y_1, y_2 \right) 
   \phi_0\left(y_2 - y_{\ell}^{(\beta)}\right) \;. 
   \label{eq:A} 
\end{eqnarray} 
Note that the definition (\ref{eq:A}) implies that the diagonal elements of 
the matrix ${\cal A}$ are proportional to the corresponding partial widths, 
${\cal A}_{11} \sim \Gamma_1$, ${\cal A}_{22} \sim \Gamma_2$.  A 
straightforward evaluation of the integrals in Eq. 
(\ref{eq:g_distribution_A}) using the substitution 
\begin{equation} 
   x = \left[a_0^{(1)}\right]^2 - 
\frac{2 m_*^2 kT G}{\pi e^2 \hbar^3 \rho_0}
\end{equation} 
yields 
\begin{eqnarray} 
   \langle G^n\rangle & = & \int dG \ G^n \  P\left(G\right) 
\end{eqnarray} 
where the distribution is 
\begin{eqnarray} 
   P\left(G\right) & = & \frac{1}{\sqrt{G}}  
   \exp\left( - \frac{1}{2} {\rm Tr}{\cal A} G\right) 
   \nonumber \\ & \times & 
   \int_0^\infty \!\! dx 
   \frac{G + x}{x^{3/2}} 
   \exp\left[ - \frac{1}{2} {\cal A}_{11} x -  
   \frac{1}{2} {\cal A}_{22} \frac{G^2}{x} \right] 
   \cosh\left[{\cal A}_{12} G \left(\sqrt{\frac{x}{G}} + \frac{G}{x}\right) 
   \right] \;. 
   \label{eq:g_distribution_general} 
\end{eqnarray} 
Note that it is only the term involving ${\cal A}_{12}$ which makes the 
remaining integral non-Gaussian and so hard to perform. However, this term is 
semiclassically small:  from Eq. (\ref{eq:A}) it follows that the leading 
semiclassical term in the off-diagonal part of the matrix ${\cal A}$ is of 
next order in $\hbar$ compared to the leading diagonal terms.  The 
$x$-integral in Eq. (\ref{eq:g_distribution_general}) is therefore dominated 
by the interval between $1/\left(G {\cal A}_{11}\right)$ and $G {\cal 
A}_{22}$, where the off-diagonal matrix element  ${\cal A}$ makes only a 
small correction {\it quadratic} in ${\cal A}_{12}$. Such a correction 
corresponds, however, to higher-order terms in $\hbar$. Corrections of this 
order were already neglected in the original semiclassical expansion of the 
Green function, and so to be consistent we discard all effects of the 
off-diagonal matrix element ${\cal A}_{12}$ here. The integral in 
Eq. (\ref{eq:g_distribution_general}) can now be easily performed. 
 
The semiclassical approximation to the conductance distribution is, then, 
simply a Porter-Thomas distribution, even in the general case:  
\begin{eqnarray} 
   P\left(G\right) & = &  
   {\left( \frac{2 \pi}{\overline{G}} \right)}^{1/2} \frac{1}{\sqrt{G}} 
   \exp\left[ - \frac{G}{2 \overline{G} } \right] \;, 
   \label{eq:g_distribution_diagonal} 
\end{eqnarray} 
where the ``local average'' conductance $\overline{G}$ is  
\begin{eqnarray} 
   \overline{G} & \equiv & \left[ \frac{1}{ \sqrt{\overline{G}_1} } 
   + \frac{1}{\sqrt{\overline{G}_2}} 
   \right]^{-2} \;,
   \label{eq:g_local_average_mixing} 
\end{eqnarray} 
and the ``partial conductance'' $G_\alpha$ is related to the partial width  
$\Gamma_\alpha$ via the standard relation 
\begin{eqnarray} 
   G_\alpha & \equiv & (\pi e^2/2 h kT) \Gamma_\alpha \;. 
   \label{eq:g_gamma} 
\end{eqnarray} 
As the semiclassical Green function, $G_{\rm sc}$, and, consequently, the 
correlator, $\partial_nC$, can be expressed as a sum of the contributions of 
``zero-length'' and longer classical trajectories, similar decompositions 
hold for the average partial width and average partial conductance: 
\begin{eqnarray} 
   \overline{\Gamma}_\alpha & = & \overline{\Gamma}^0_\alpha + 
   \overline{\Gamma}^{\rm osc}_\alpha \\ 
   \overline{G}_\alpha & = & \overline{G}^0_\alpha + 
   \overline{G}^{\rm osc}_\alpha  
\end{eqnarray} 
where the ``oscillatory'' parts, $\overline{\Gamma}^{\rm osc}$ and 
$\overline{G}^{\rm osc}$, depend on the longer classical trajectories and are 
of next order in $\hbar$ compared to the smooth  contributions, 
$\overline{\Gamma}^{0}$ and $\overline{G}^{0}$ which are the zero-length 
contributions. A consistent semiclassical approximation, as in Eq. 
(\ref{eq:Gsc}), then requires expanding $\overline{G}$ and keeping only the 
linear terms in the oscillatory contribution. We thus obtain 
\begin{eqnarray} 
   \overline{G} & = &  
   \frac{\overline{G}_1^{0}\overline{G}_2^{0}} 
   {\overline{G}_1^{0}+\overline{G}_2^{0} 
   +2 \sqrt{\overline{G}_1^{0}\overline{G}_2^{0}}} 
   \left[1 +  
   \frac{1}{ 1 + \sqrt{\overline{G}_1^0 /\overline{G}_2^0 }} 
   \frac{\overline{G}_1^{\rm osc}}{\overline{G}_1^0 } 
   +  
   \frac{1}{ 1 + \sqrt{\overline{G}_2^0 /\overline{G}_1^0 }} 
   \frac{\overline{G}_2^{\rm osc}}{\overline{G}_2^0 } 
   \right] \;. 
   \label{eq:g_local_average} 
\end{eqnarray} 
 
We now proceed to the semiclassical calculation of the ``local average''
partial width $\overline{\Gamma}$. The defining equation 
(\ref{eq:gamma_local_average}) involves the functional inverse of 
the Green function, which is a hard object to calculate. Instead, we 
will use the original definition (\ref{eq:gamma_partial}), which for the
local average partial width yields
\begin{eqnarray}
   \overline{\Gamma} & = & 
   \frac{2 \pi \hbar^4}{m_*^2}   
   \sum_{\alpha} 
   \rho_0^{(\alpha)} 
   \int d{y_1} \  \int d{y_2} \ 
   \phi_\alpha\left(y_1 - y_\ell\right) 
   \phi_\alpha\left(y_2 - y_\ell\right)^*      
   \partial_nC\left(y_1,y_2\right)
   \label{eq:gamma_partial_average} 
\end{eqnarray} 
where the correlation function of the normal derivatives of the dot
wave functions $\partial_nC\left(y_1,y_2\right)$ is related to the
semiclassical Green function by Eq. (\ref{eq:dCsc}).

If we now use some information about the lead wave functions, we can obtain an 
explicit expression for the average width $\overline{\Gamma}$ in terms of the 
classical dynamics in the dot.  When, as we assumed above, the tunneling from 
the lead to the dot is dominated by the lowest transverse energy subband in 
the constriction between the lead and the dot\cite{nato-book-2}, the 
transverse potential in the tunneling region can be taken quadratic: 
$U_{\ell} \sim \kappa \left(y-y_\ell\right)^2$.  In this case, the transverse 
dependence of the lead wave function is simply a harmonic oscillator 
wave function, so that at the edge of the dot $\phi_0 \simeq c_\ell \exp [ - 
(y - y_\ell )^2/2 a_{\rm eff}^2 ]$, where  $y_\ell$ is the center of the lead 
and constriction, and the effective width is $a_{\rm eff} = 
\sqrt{\hbar}/\sqrt[4]{2 \kappa m^*}$.  While the exact form of the lead 
wave function is not crucial, the $\hbar$-dependence of the width is important 
for the semiclassical argument which follows; note that $a_{\rm eff} \sim 
\sqrt{\hbar}$ does not depend on a particular transverse potential. 
 
Using this information about $\phi_0$ in the expression for the diagonal 
matrix elements ${\cal A}_{11}$, ${\cal A}_{22}$, 
we see that the lead wave function restricts the integration 
to a semiclassically narrow region of width $a_{\rm eff} \sim \sqrt{\hbar}$. 
This allows one to express the contribution of the open trajectories 
entering the Green function in terms of an expansion near their closed 
neighbors:
\begin{eqnarray}
\overline{\Gamma} & = & \overline{\Gamma}_0 +  
\frac{16}{{m^*}} \int dy \int dp_y 
f_W\left(y, p_y\right)
\sum_\alpha
\sqrt{ 
\frac{(p^\alpha_i)_n (p^\alpha_f)_n}
     { m_{11}^\alpha + m_{22}^\alpha + 2 }
}
\exp\left[ - \frac{i}{\hbar} 
\frac{2 m_{12}^\alpha}{m_{11}^\alpha + m_{22}^\alpha + 2} 
\left(p_y - \bar{p}^\alpha_y\right)^2 \right] \nonumber \\
& \times & 
\exp\left[i \frac{S_\alpha\left(y,0;y,0;\varepsilon\right)}{\hbar} \right]
\label{eq:closed}
\end{eqnarray}
where $\overline{\Gamma}_0$ is the monotonic part of the resonance width, 
 $(p_i)_n$ and $(p_f)_n$ are the normal components of the initial 
and the final momenta of the closed orbit $\alpha$, the momentum
$\overline{\bf p} \equiv \left({\bf p}_i + {\bf p}_f\right)/2$, the 
$2\times2$ monodromy matrix\cite{Gutzwiller}  
$M_\alpha \equiv (m^{\alpha}_{ij})$ is 
defined via the  linearization
of the Poincar\'e map near the closed orbit $\alpha$ , and  
calculated at the contact point near the lead. 
We have also introduced in Eq. (\ref{eq:closed})
the Wigner transform $f_W^e$ of the lead wave function
 \begin{eqnarray}
f_W^e\left(y, p_y\right) \equiv h^{-1} \int d \Delta y
\ \phi_0\left(y - \Delta y/2, 0 \right)
\ \phi_0^*\left(y + \Delta y/2, 0 \right)
\ \exp\left( i p_y \Delta y/ \hbar \right)
\label{define_wigner}
\end{eqnarray}
which describes the distribution in transverse 
position and momentum of electrons tunneling into the dot.

In leading order in the distance between the contact point
$y$ of the closed orbit $\alpha$, and the center of the lead $y_\ell$,
the action of the closed orbit $S_\alpha$ scales linearly:
\begin{eqnarray}
S_\alpha\left(y,0;y,0\right) \propto S\left(y_\ell,0;y_\ell,0\right)
+ \Delta{p}_y^{\alpha} \left( y - y_\ell\right)
\end{eqnarray}
where $\Delta p_y$ is the change of transverse momentum 
after the traversal of the closed orbit. Assuming e.g. a Gaussian
form of the lead wave function,  the contribution of 
each of these closed orbits is suppressed by a factor exponentially small in 
$\Delta p_y^2$. This suppression is the effect of the mismatch of the 
closed orbit (momentum) with the distribution of transverse momentum at the 
lead, which is centered at zero with width $\delta p_\ell \sim \hbar/a_{\rm 
eff} \sim \sqrt{\hbar}$ for the lowest subband. Therefore, only closed 
orbits with {\it semiclassically} small momentum change $\Delta p$ 
contribute to the width. This in turn implies that the closed orbit is 
located semiclassically close (within a distance $\sim \sqrt{\hbar}$) to a 
{\it periodic orbit} for which $\Delta p \equiv 0$. 
Using this proximity to a periodic orbit we can re-express the
actions and momenta of the injection orbits in terms of the properties
of their periodic neighbors (labeled by the index $\mu$) as follows:
\begin{eqnarray}
S_\alpha\left(y, 0; y, 0\right) & \simeq & S_\mu 
+ \frac{ {\rm Tr}\left[M_\mu\right] - 2}{2 m_{12}^\mu}
\left(y - y_\mu\right)^2, \label{eq:S:clo-po} \\
\bar{p}_y^\alpha & \simeq & p^\mu_y + 
\frac{m^\mu_{11} - m^\mu_{22}}{2 m_{12}^\mu} \left(y - y_\mu\right)
\label{eq:p:clo-po}
\end{eqnarray}
Substitution of (\ref{eq:S:clo-po}) and (\ref{eq:p:clo-po}) into Eq.
(\ref{eq:closed}) and integration over $y$ yields\cite{Nar_cbdyn1} 
\begin{eqnarray} 
   \overline{\Gamma} = \overline{\Gamma}_0 +   
   \sum_{\mu:{\rm p.o.}} A_\mu  
   \cos\left( \frac{S_\mu}{\hbar} + \phi_\mu \right) 
   \label{eq:gamma_sc} 
\end{eqnarray} 
where the monotonic part is 
\begin{eqnarray} 
   \overline{\Gamma}_0 = \case{ \sqrt{\pi}}{2} c_\ell^2 a_{\rm eff} 
   \frac{p^2}{m^*} \ 
   e^{-\zeta} \ 
   \left[I_0\left(\zeta\right) 
   + I_1\left(\zeta\right) 
   \right], \ 
   \ \zeta = \frac{p^2 a_{\rm eff}^2}{2 \hbar^2} , 
\end{eqnarray} 
the amplitude is 
\begin{eqnarray} 
   A_\mu & = & 4 \sqrt{2} \ \frac{\hbar c_\ell^2 p^\mu_z}{m^*} 
   \ \left[{\rm Tr}^2\left[M_\mu\right] 
   \left(1 + \sigma_+^2 \right) 
   \left( 1 + \sigma_-^2 \right) \right]^{-1/4} 
   \nonumber \\ 
   & \times &  
   \exp 
   \left(  
   -  
   \frac{\sigma_+^2 \bar{p}^2} 
        {\left(1 + \sigma_+^2\right)}  
   -  
   \frac{\sigma_-^2 \bar{y}^2} 
        {\left(1 + \sigma_-^2\right) }  
   \right) 
   \label{eq:gamma_osc_amp} 
\end{eqnarray} 
with 
\begin{eqnarray} 
   \sigma_\pm  & \equiv & \case{1}{2} \left[ 
   \overline{m}_{12} \! - \! \overline{m}_{21} \pm    
   \sqrt{\left(\overline{m}_{22}\! -\! \overline{m}_{11}\right)^2 
   + \left(\overline{m}_{21}\! +\! \overline{m}_{12}\right)^2} \right] 
   \nonumber  
   \\ 
   \overline{m}_{ij} & \equiv &  
   \frac{2 m^\mu_{ij}} 
   { {\rm Tr}\left[M_\mu\right] + 2} 
   \left( \frac{a_{\rm eff}^2}{\hbar} \right)^{\frac{j-i}{2}} 
   \nonumber \\ 
   \theta & \equiv & \case{1}{2} \arctan\left( 
   \frac{\overline{m}_{22} - \overline{m}_{11}} 
   {\overline{m}_{21} + \overline{m}_{12}} 
   \right) 
   \\ 
   \bar{y} & \equiv & 
   \cos\theta \left(y_\mu - y_\ell\right)/a_{\rm eff} 
   + \sin\theta \ p^\mu_y a_{\rm eff}/\hbar 
   \nonumber \\ 
   \bar{p} & \equiv  & 
   \cos\theta \ p^\mu_y a_{\rm eff}/\hbar 
   - \sin\theta \left(y_\mu - y_\ell\right)/a_{\rm eff} \, , 
   \nonumber  
\end{eqnarray} 
and $\phi_\mu$ is a slowly varying phase. Here $I_n$ is the Bessel function 
of complex argument, $p$ is the magnitude of the electron momentum, ${\bf 
p}^\mu$ is the electron momentum for the periodic orbit $\mu$ at the bounce 
point (turning point), 
$y_\mu$ is the bounce point coordinate, $S_\mu$ is the action of the 
periodic orbit, and $M_\mu \equiv \left(m_{ij}^\mu\right)$ is the 
corresponding monodromy matrix\cite{Gutzwiller}.  Note the sharp suppression 
of the oscillatory effects in Eq. (\ref{eq:gamma_osc_amp}) if the periodic 
orbit does not match up to the lead wave function in {\it both} position and 
momentum space. The mismatch is characterized by $\overline{y}$ and 
$\overline{p}$; the most favorable case is that of a perpendicular periodic 
orbit hitting the edge of the dot right at the center of the lead, $p_y^\mu 
\!=\! 0$ and $y_\mu \!=\! y_{\ell}$ so that $\overline{y} \!=\! 
\overline{p} \!=\! 0$. 
 
An explicit expression for the average conductance follows from the relation 
between the partial width and the partial conductance 
Eq.~(\ref{eq:g_gamma}).  Using Eq.~(\ref{eq:g_local_average}), we see that 
$\overline{G}$ can be written in the form 
\begin{eqnarray} 
   \overline{G} = \overline{G}_0 +   
   \sum_{\mu:{\rm p.o.}} B_\mu  
   \cos\left( \frac{S_\mu}{\hbar} + \phi_\mu \right) 
   \label{eq:g_sc} 
\end{eqnarray} 
where $B_\mu$ is simply related to $A_\mu$, $\overline{G}_1^0$, and 
$\overline{G}_2^0$.  This together with 
Eq.~(\ref{eq:g_distribution_diagonal}) defines both the average conductance 
and its fluctuations. 
 
Further characterization of the peak fluctuations can be obtained from the 
peak-to-peak correlation function: this is a particularly interesting 
quantity because of the correlations sometimes observed 
experimentally~\cite{Chang96,MarcusFolk96}, as discussed in the Introduction.   
A natural measure of the statistics of nearby peaks is given by  
$ \delta G\left(E_m\right) \!\equiv\! G\left(E_m\right) \!-\!  
\langle G\left( E_n\right) \rangle_n $ 
in terms of which the correlation function is 
\begin{eqnarray} 
   {\rm Corr}_m\left[\delta G, \delta G\right] = 
   \langle \delta G\left(E_{n+m}\right) \delta G\left(E_n\right)  
   \rangle_n / 
   \langle [ \delta G(E_{n}) ]^2 \rangle_n \;. 
   \label{eq:correlator_define} 
\end{eqnarray} 
Substituting the conductance distribution (\ref{eq:g_distribution_diagonal}) 
into (\ref{eq:correlator_define}), we obtain 
\begin{eqnarray} 
   {\rm Corr}_m & = &  
   \delta_{m,0} + \left(1 - \delta_{m,0}\right)\times 
   \frac{  
   \sum_\mu B_\mu^2 
   \cos\left( \frac{\tau_\mu \Delta}{\hbar} m \right)}
   {4 \bar{G}_0^2 + 3 \sum_\mu B_\mu^2} 
   \, . 
   \label{eq:sc_correlator} 
\end{eqnarray} 
 
Throughout this paper we have concentrated on energy (or equivalently peak
number) as the tuning parameter causing the peak height variation.  This is
just an example: exactly analogous considerations apply to any parameter
causing changes in the wave functions of the quantum dot.  In particular,
similar oscillatory behavior is expected in the height of a {\it given} peak
as a function of magnetic field, often the most experimentally accessible
parameter.  As the field varies, the change in the action of a periodic orbit
is proportional to the (directed) area that it encloses.  Thus, the peak
heights should exhibit an oscillatory envelope whose frequencies are
proportional to the areas of the periodic orbits.

\begin{figure}[t] 
\begin{center} 
\leavevmode 
\epsfxsize = 11.9 cm 
\epsfbox{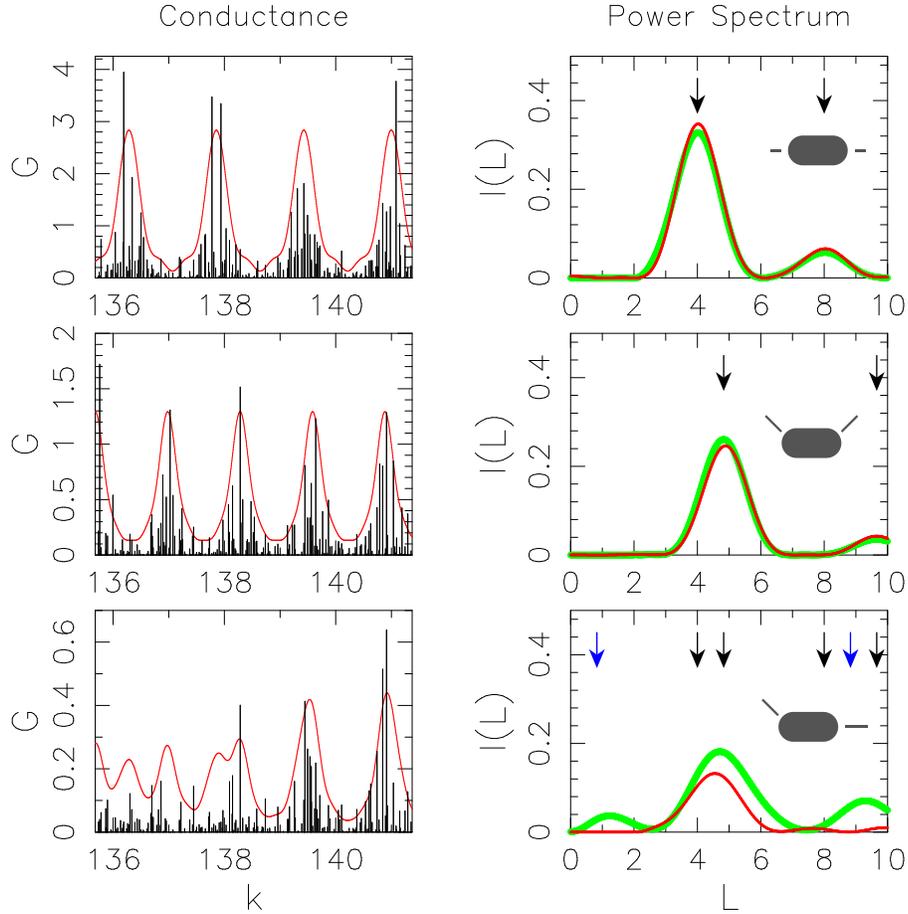} 
\end{center} 
\caption{ 
The peak conductances (left column) from tunneling through  
subsequent energy levels 
in the stadium quantum dot and the corresponding ``length spectra'' 
$I\left(L\right)$ 
(right column) for different lead configurations (shown in the insets). 
In the peak conductance plots, each peak is placed at  
the wavevector $k$ corresponding to its level; $R$ is the radius of 
the half-circle parts of the stadium dot. A Gaussian lead 
wave function appropriate for tunneling from a single transverse mode 
is used with width $ka_{\rm eff} = 15$. The red curves represent the  
semiclassical envelopes, defined by the contributions of the relevant periodic 
orbits (top: ``diameter'' orbit, middle: ``$V$''-shaped orbit,  
bottom: both diameter- and $V$-orbits). 
Length spectrum of the oscillations in $G(k)$ obtained from the Fourier 
power, numerical (green) and semiclassical (red) results are compared. The  
power is normalized to the mean conductance. The arrows at the top show
the positions of the relevant periodic orbits and their repetitions 
(black arrows), as well as the ``combination lengths'' $L_1 \pm L_2$ (blue 
arrows). 
In the top panel, the peak at 
$L/R \simeq 4$ is the diameter, that at $8$ is its repetition. In the  
middle panel, the peak at  
$L/R = 2\protect\left(1 + \protect\sqrt{2}\protect\right)  
\protect\approx 4.8$ 
corresponds to the $V$-shaped orbit, the peak at $L/R \approx 9$ represents 
its repetition. In the bottom panel (asymmetric leads), the broad peak 
at $L/R \simeq 4.5$ represents the total contribution of both diameter- 
and $V$-shaped orbits. 
For the stadium dot,  the principal peaks appear at $4$ and $4.8$,  
because we use only the wave functions symmetric 
about the vertical and horizontal symmetry axes (equivalent to using only 
the even-even states of the stadium). 
Note the excellent agreement between the semiclassical theory and the 
numerical results for symmetric leads, and adequate representation of the 
principal peak for asymmetric leads. 
} 
\label{fig_big} 
\end{figure}

\section{Comparison with numerics and experiment} 
\label{numerics} 
 
Since one of the main theoretical results of the present paper concerns the
periodic modulation of the Coulomb blockade peak heights, it is natural to
consider the Fourier power spectrum of $G_{\rm peak}(k)$. In Fig.
\ref{fig_big} we present a comparison of the numerical and semiclassical
power spectra, calculated for a chaotic (stadium) dot, for three different
placements of the leads.  Previously, we reported the case for leads
placed symmetrically as in the upper plot of Fig. \ref{fig_big} for $kR = 70$.  
The exact conductance peaks are obtained numerically
from Eq.~(\ref{eq:g_max}) with the eigenstates being constructed using the
the method of Ref.~\cite{Julie}. To observe the variation in peak height,
we vary the energy, or equivalently the wavevector $k = p/\hbar$, which
changes the number of electrons on the dot as more levels are filled.
The data clearly demonstrate that the power spectrum has well-defined 
peaks corresponding to periodic orbits. The numerical results for the 
symmetric leads show excellent agreement with the semiclassical prediction.
 
The situation is however different for asymmetrically positioned leads when
there is no single short periodic orbit connecting both leads. In this case,
only the main peak corresponding to the first repetition of the relevant
periodic orbits, the ``diameter'' and the $V$-shaped orbit, is adequately
reproduced. The higher-frequency behavior, however, is substantially
different from the semiclassical prediction. We attribute this difference to
the non-linear mixing of the oscillations of different partial widths,
neglected in our derivation of Eq.~(\ref{eq:g_local_average}). The 
pronounced peak at the difference length $L_V - L_D$, where $L_V$ and $L_D$ 
correspondingly represent the lengths of the $V$-shaped and diameter orbits, 
strongly indicates that, although semiclassically small, the mixing effects 
of higher order terms in Eq.~(\ref{eq:g_local_average_mixing}) can be 
significant in the experimentally relevant parameter range.  We numerically
verified that the sum and difference lengths can be partially obtained by 
Eq.~(\ref{eq:g_local_average_mixing}).

\begin{figure}[t] 
\begin{center} 
\leavevmode 
\epsfxsize = 7.0cm 
\epsfbox{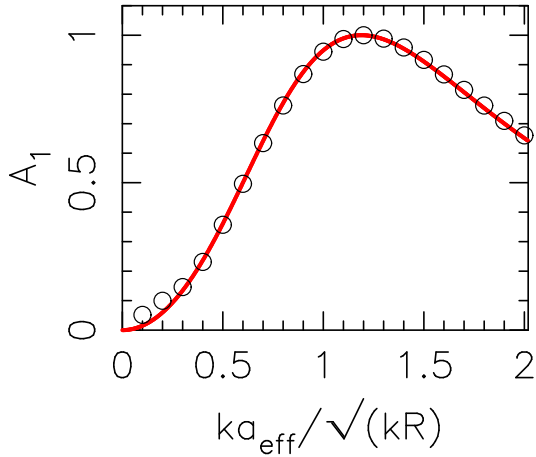} 
\end{center} 
\caption{ 
The dependence of amplitude of the length spectrum peak at  
$L/R \simeq 4$ on $ka_{\rm eff}$ for $kR = 70$. The leads are 
 symmetrically attached to the middle of the 
semicircle segments of the stadium dot.} 
\label{fig_A1} 
\end{figure}

As follows from Eq. (\ref{eq:gamma_sc}), the oscillatory component of the
``local average'' conductance and the height of the corresponding peak in the
power spectrum depends nontrivially on the position and the width of the
lead. This dependence is illustrated in Fig. \ref{fig_A1}, where we plot the
amplitude of the ``diameter'' orbit contribution to the conductance as a
function of $k a_{\rm eff}$ extracted from numerical length spectrum and the
corresponding semiclassical prediction.

\begin{figure}[t] 
\begin{center} 
\leavevmode 
\epsfxsize = 14.0cm 
\epsfbox{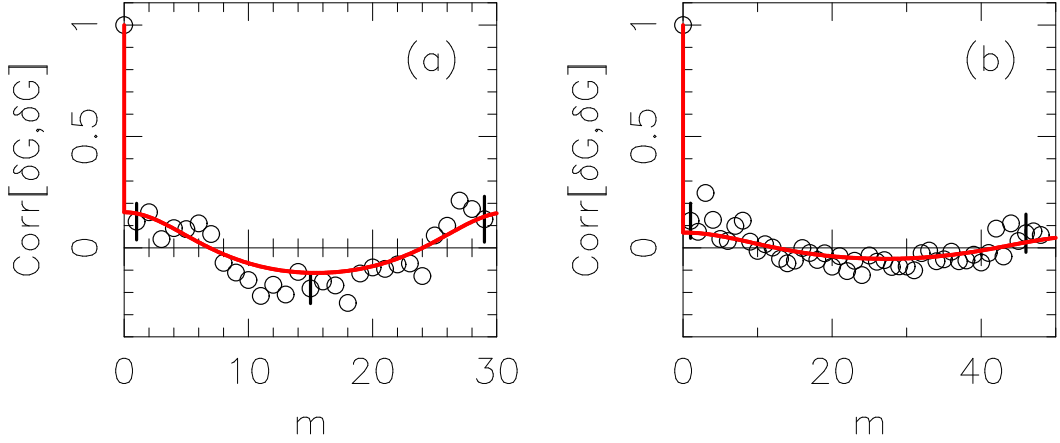} 
\end{center} 
\caption{The peak-to-peak conductance correlation function 
for (a) symmetrically placed leads (attached to the ``diameter'' of the 
stadium dot), and (b) asymmetric leads (as in the inset to the length 
spectrum at the bottom of Fig. \protect\ref{fig_big}.
The numerical correlation function  
(circles with typical error bars)---the average of all pairs of  
peaks $m$ peaks apart---is in good agreement with the semiclassical  
theory (red).  
The agreement for small $m$ is surprising 
since this regime is not semiclassical, but shows how dynamics can give 
rise to correlations even between  
nearest-neighbors. The difference between the periods of the modulation  
in (a) and (b) is accounted for by the difference in the values of $kR$ 
used for the correlation function: 
the calculation for (a) is performed 
near $kR = 70$, while (b) corresponds to an interval near 
$kR = 140$.  
} 
\label{fig_corr0} 
\end{figure}

In Fig. \ref{fig_corr0} we compare the semiclassical correlation function
with numerical data for the stadium dot. The oscillatory behavior for large
separations reflects the peak in the corresponding power spectrum in Fig.
\ref{fig_big} and is in agreement with the semiclassical result.  The
positive correlation for nearest neighbors is also in agreement with the
semiclassical theory, demonstrating the influence of dynamics even in this
apparently non-semiclassical regime.

\begin{figure}[t] 
\begin{center} 
\leavevmode 
\epsfxsize = 8.6cm 
\epsfbox{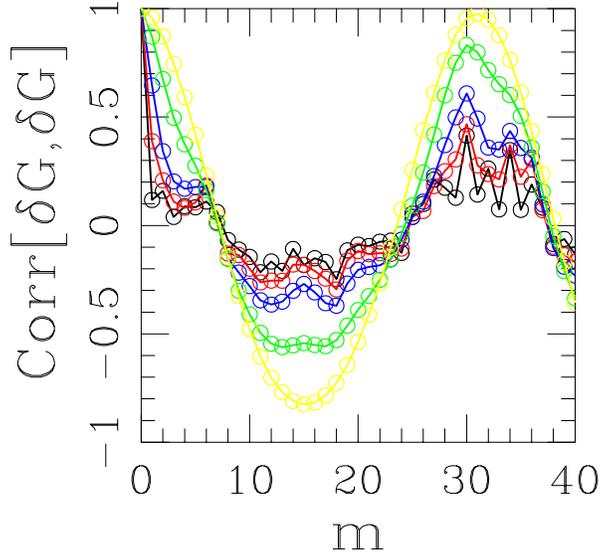} 
\end{center} 
\caption{ 
The peak amplitude correlation function for stadium-shaped quantum dot with  
symmetrically attached leads, for the temperature  
$T = 0.25 \Delta ({\rm red \ curve}), 0.5 \Delta ({\rm blue}),  
\Delta ({\rm green}), 2\Delta  ({\rm yellow})$ and $kR = 70$.    
} 
\label{fig_corrT} 
\end{figure}

When $T \! \gg \! \Delta$, the major source of correlations between
neighboring peaks is the joint contribution of several resonances to the same
conductance peak\cite{Alhassid98}. In this regime the ``nearest-neighbor''
correlator is ${\rm Corr}_{m = 1} \! \sim \! 1$, and the dynamical effect
accounts for only a small correction to the correlation function. However,
for low temperature $T \! \leq \! \Delta$, the correlations due to
temperature are exponentially suppressed. In this regime, as illustrated in
Fig. \ref{fig_corrT}, the correlations induced by dynamical modulation
dominate, and they account for the experimentally observed enhancement of
correlations at low temperatures\cite{MarcusPatel98}.  For finite temperature
each resonance is weighted by combinations of Fermi-Dirac functions and
occupation numbers~\cite{beenakker}.  The occupation numbers used in Fig.
\ref{fig_corrT} were obtained by employing a recursion
relation~\cite{Brack}; see Appendix~\ref{append}.  As the temperature 
increases more resonances contribute to a single conductance peak, and thus 
dampening the effects of the longer orbits.

\begin{figure}[t] 
\begin{center} 
\leavevmode 
\epsfxsize = 14.0cm 
\epsfbox{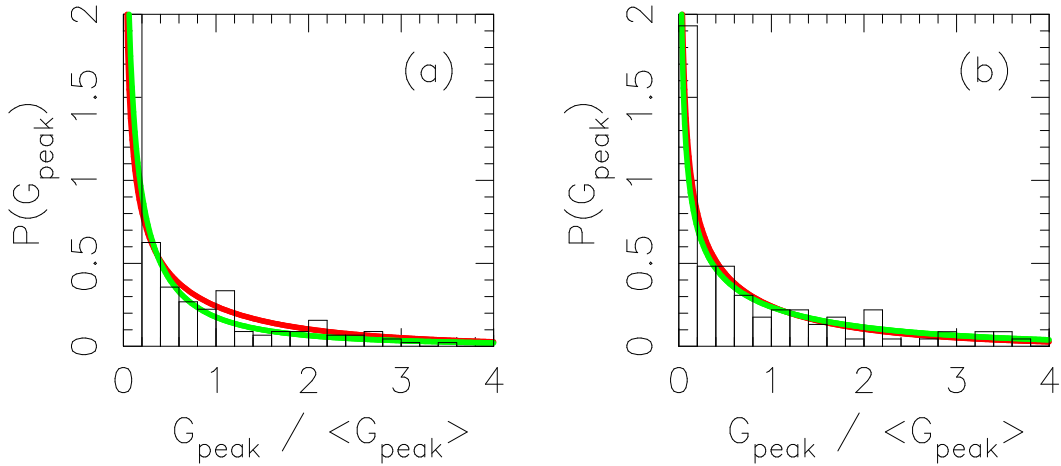} 
\end{center} 
\caption{ 
Conductance statistics:  probability distribution function for  
(a) symmetric leads at $kR = 140$ and (b) asymmetrically placed leads
at $kR = 70$.  The numerical 
probability distribution (histogram) is for the entire range of data in 
Fig. \protect\ref{fig_big} and is compared to both  
the semiclassical theory (red) and the 
standard statistical theory based on random wave functions (green). 
The two theories predict nearly the same result for this quantity 
(especially for asymmetric leads, where the dynamical modulation is  
weaker), and both are consistent with the numerics. 
} 
\label{fig_distr} 
\end{figure}

In Fig. \ref{fig_distr} we present the results of the calculation of the
probability distribution of $G_{\rm peak}$ for a stadium quantum dot for both
the ``symmetric'' and ``asymmetric'' placement of the leads.  For comparison,
we show both the actual distribution, Eqs.
(\ref{eq:g_distribution_diagonal}) (\ref{eq:g_local_average}), and the
standard Porter-Thomas result without any account of the modulation of the
average conductance:  $P(G_{\rm peak}) = \sqrt{4/\pi G_{\rm peak}}
\exp(-G_{\rm peak})$. As the {\it individual} peak-height distribution is
essentially a local measure, it is not strongly sensitive to the
correlations, and both the standard and the dynamical theories predict nearly
the same result, and both are consistent with numerical calculation.  This
explains why no dynamical effect was observed in the experimental peak-height
probability distribution\cite{Chang96,MarcusFolk96}.

In contrast, the periodic modulation of the peak heights has been observed in
several recent
experiments\cite{MarcusFolk96,MarcusPatel98,MarcusCronenwett97}. The clearest
observation is in Ref. \onlinecite{MarcusPatel98}: the data in their Fig.
\ref{fig_big} show modulated peak heights as a function of the number of
electrons in the dot. In their trace of 90 peaks, approximately six
oscillations are visible, yielding a period of $\sim \! 15$ peaks. In our
treatment, this period is related to the period of fundamental oscillation in
Eq. (\ref{eq:gamma_sc}).  A variation in action $\Delta S_\mu$ can arise from
changes in either or both the Fermi momentum and the dynamics, i.e.~the
lengths of the orbits.  If only the momentum varies, then the fundamental
period is given by $\left(\frac{1}{h}\frac{\partial S_\mu}{\partial
\varepsilon} \right)^{-1} \equiv h/\tau_\mu$ where $\tau_\mu$ is the period
of the relevant orbit, and the ratio of this to the level spacing $\Delta$
gives the period of the peak heights.  In the billiard
approximation, $\tau_{\mu} \equiv L_\mu / v_F$, where $L_\mu$ is the length
of the periodic orbit and $v_F$ is the Fermi velocity, which can be
calculated from the experimental density\cite{Marcus_comm}. Using the
appropriate spin-resolved level spacing $\Delta \simeq 10$ $\mu$eV (which is
half of the spin-full value from the measurements in Ref.
\onlinecite{MarcusPatel98}), the experimental period implies $L_\mu \simeq
4.5 {\mu}m$.  This value is inconsistent with the period given by the
shortest periodic orbit contacting the leads \cite{correction}---the orbit
from the lead to the ``pin'' gate and back whose length we estimate to be
$0.9 \mu$m.

Changes in the dynamics should also be considered.  By examining the
configuration of the dot in the insert of Fig. 1 of
Ref.~\cite{MarcusPatel98}, the gate voltage appears to be  situated on the
shortest periodic orbit of each lead.  By making the gate voltage more
negative the electron will have a shorter path and thus contribute to the
change of the action in Eq.~(\ref{eq:gamma_sc}).  If the Fermi energy of the
system remains constant, then we can calculate a plausible range for the
period using two extreme simplified models for the deformation of the
boundary.  First, the gate voltage is modeled as a small local semicircular
deformation.  Equating the area of the semicircle to the number of peaks
times the change of area caused by adding one electron on the dot without a
change in the Fermi energy, we estimate the period to be $\sim \! 3$ peaks
independent of the Fermi energy.  The other extreme is considering the entire
side to move uniformly.  The same procedure yields a period proportional to
the square root of the number of electrons on the dot.  Assuming that the
typical dot in these experiments has 100 electrons~\cite{MarcusPatel98}, we
obtain a period of $\sim \! 13$ peaks.  Thus, the experimental result of a
period of 15 peaks cannot be obtained solely by the orbit length change,
but must also include a change due to the difference in Fermi energies
of successive peaks.  A more detailed model of the
gate voltage is necessary in order to make a better prediction of the
oscillation period.

A similar approach to the peak modulation as a function of magnetic field is
also in agreement with the experimental
results\cite{MarcusFolk96,MarcusCronenwett97}, where a quasi-periodic
modulation of the peak heights was observed with the period ${\Delta}B \simeq
35$mT.  In our treatment, this period is given by the ratio of flux quantum
$hc/e$ to the area ${\cal A}_0$ enclosed by the periodic orbit. From the
experimental oscillation, we obtain ${\cal A}_0 \simeq 0.12 {\mu}$m$^2$. This
is consistent with the total area of the dot, $0.32
{\mu}$m$^2$\cite{MarcusFolk96}, considering that there is likely to be some
cancellation of fluxes between different parts of the orbit.

A puzzling feature of the initial experiments was that the dynamical
modulation of the Coulomb Blockade peak heights was not seen in the
experiment of Ref. \onlinecite{Chang96}. We attribute this behavior to two
factors: the positioning of the leads relative to the gate and a relatively
small mean free path.  First, if the gate used to change the number of
electrons is not along the shortest periodic orbit of either lead and the
Fermi energy does not change appreciably in the dot, then one should not
observe oscillations in the conductance peaks. In the geometry of Ref.
\onlinecite{Chang96} the leads and gate seem to be rather disconnected so
this may be a factor.  Second, in this experiment the mean free path $\ell
\sim 0.4{\mu}$m only marginally exceeds the typical size of the dot $d \simeq
0.25{\mu}$m, while the length of the shortest periodic orbit is at least
twice the effective ``diameter'' $d$ of the dot: $L_{\rm min} > 2d \simeq 0.5
{\mu}m > \ell$.  If the mean free path is caused by short-range diffractive
scattering, the dynamical effects are suppressed and will not affect the
Coulomb Blockade measurements.  However, in the opposite limit of a smooth
scattering potential, dynamical effects caused by coherent branched flow
\cite{Tomsovic00} may still be present. The short mean free path measured in
Ref.  \onlinecite{Chang96} suggests the presence of impurities in the
two-dimensional electron gas layer leading to a short-range scattering
potential and so suppression of dynamical effects.

The agreement of our semiclassical theory with experiment may seem
surprising, since the adding of electrons  changes the effective potential
defining the dot because of the added charge.  However, experiments  on
``magnetofingerprints'' of the peaks\cite{StewartMarcus} suggest robustness
of the effective potential---its change from peak to peak seems to be small
in this case.  In contrast, to affect the dynamical modulation one must
substantially change the action of the shortest periodic orbit, which
typically requires a much larger change in potential such as could be
caused by the external gates.

\section{Summary} 
\label{summary} 
 
In conclusion, using semiclassical methods, we developed a dynamical
statistical  theory of Coulomb blockade peak heights in chaotic quantum dots.
We derived the peak height distributions and the correlation functions, and
showed that the corrections to the corresponding results of the standard
statistical theory can be expressed in terms of the classical periodic orbits
of the dot. Both our analytical results and numerical simulations clearly
demonstrate that the dynamical effect is significant for both symmetric and
asymmetric lead placements.

We close with two further experiments suggested by our results. First, if the
tuning parameter used to change the number of electrons, such as a gate
voltage, does not change the action of the dominant periodic orbit, then no
modulation connected to that orbit should be seen. In particular, gates which
affect different parts of the dot may produce different oscillatory behavior.
Second, several samples made in a robust geometry---a circle with directly
opposite leads, for example---should show the same modulation. Any deviations
from the same behavior would be a sensitive indication of the material
quality.

\section*{Acknowledgments} 
We gratefully acknowledge stadium eigenfunction calculations 
by J.~H.~Lefebvre, and helpful discussions with C. M. Marcus and 
M.~Srednicki. We thank L. Kaplan for making available  
Ref.~\onlinecite{Kaplan} which we received during the final stages of 
this work. 

\appendix
\section{Temperature Calculations}
\label{append}

For nonzero temperatures the conductance is obtained from a weighted
sum over the zero temperature partial widths 
$\Gamma_\lambda$~\cite{beenakker}.  For symmetric leads this yields
\begin{equation}
  G = {e^2 \over h} {\pi \over 4 k T} \sum_\lambda w_\lambda \Gamma_\lambda 
\end{equation}
If $kT, \Delta \ll e^2/C$, then the weights are given by
\begin{equation}
  \label{wlambda}
  w_\lambda = 4f(\Delta F_{N_0} - \tilde{E}_F) \langle n_\lambda \rangle_{N_0}
  \left[ 1 - f(E_\lambda(N) - \tilde{E}_F) \right]
\end{equation}
where $\Delta F_N$ is the change in the canonical free energy from
$N - 1$ to $N$, $\langle n_\lambda \rangle_N$ is the canonical
occupation, $\tilde{E}_F = E_N + (N - 1/2)e^2/C$ is an
effective Fermi energy and $f(\epsilon) = [1 + \exp(\epsilon/kT)]^{-1}$ 
is the Fermi-Dirac function.

To obtain the canonical free energy and canonical occupation number
we use a recurrence relation developed by Brack, Genzken and 
Hansen~\cite{Brack} for the partition function $Z(N, M; \beta)$;
$N$ is the number of particles, $M$ is the number of levels
and $\beta = 1/kT$.  The final result for the partition function 
will not numerically depend upon $M$ for large $M$.  The partition function 
is formally given by
\begin{equation}
  \label{Zdef}
  Z(N, M; \beta) \equiv 
  \sum^{I_{NM}}_{\alpha = 1} \exp (-\beta E_\alpha(N) ) =
  \exp(-\beta E_0 ) z(N, M; \beta)
\end{equation}
where
\begin{equation}
  z(N, M; \beta) \equiv
  \sum^{I_{NM}}_{\alpha = 1} \exp \Bigl(-\beta [ E_\alpha(N) - E_0] \Bigr)
\end{equation}
Here $E_\alpha(N)$ is the sum of the energy of the single particle
occupied levels $\epsilon_M$ which does not include the charging energy,
$E_0$ will be defined below, and 
$I_{NM}$ is the number of ways to fill $M$ levels with $N$ identical 
particles.  The recurrence relation derived in Ref. \onlinecite{Brack} is 
\begin{equation}
  \label{recurrance}
  Z(N, M; \beta) = Z(N, M - 1; \beta) + \exp(-\beta \epsilon_M)
  Z(N - 1, M - 1; \beta) \ \ {\mathrm for} \ \ N \ge 1, M \ge N
\end{equation}
with the constraints
\begin{eqnarray}
  Z(0, M; \beta) \equiv 1  \ \ \forall M \ge 0 \label{constraint1} \\
  Z(N, N - 1; \beta) \equiv 0 \ \ \forall N \ge 1 \label{constraint2} \;.
\end{eqnarray} 
Note that the same recurrence relation also holds for $z(N,M; \beta)$.
The choice $E_0(N) = \sum_{m=1}^N \epsilon_m$ yields the result
\begin{equation}
   \label{zNN}
   z(N,N; \beta ) = 1 \;.
\end{equation}

Using the conditions (\ref{constraint2}) and (\ref{zNN}) as a starting
point for the recurrence relation Eq. (\ref{recurrance}), we obtain
$z(N, M \to \infty ; \beta)$ and thus 
$Z(N, M \to \infty ; \beta)$.
For the small temperatures that we consider, the convergence of the
recurrence relation is rapid.

Similarly, one can calculate a modified partition function
$Z_{\lambda}' (N, M; \beta )$ which has level $\lambda$ removed from
the spectrum. The probability for level $\lambda$ to be
unoccupied, $P\{n_\lambda =0\}$, is, then, simply $Z_{\lambda}' / Z$.
In terms of this probability, the average occupation numbers are given by 
$\langle n_\lambda \rangle_{N} = 1 - P\{n_\lambda =0\}$.
Finally, the canonical free energy for $N$ electrons, $F(N)$, appearing
in Eq. (\ref{wlambda}) is
\begin{equation}
   F(N) = -{1 \over \beta} \ln Z(N, M \to \infty ; \beta) \;.
\end{equation}


\begin{references} 
 
\bibitem{nato-book} 
H.~Grabert and M.~H.~Devoret, 
{\it Single Charge Tunneling: Coulomb Blockade Phenomena in Nanostructures} 
(Plenum Press, New York, 1992). 
 
\bibitem{jalabert}  
 R.~A.~Jalabert, A.~D.~Stone, and Y.~Alhassid, 
{Phys. Rev. Lett.} {\bf 68}, 3468 (1992). 
 
\bibitem{nato-book-2}  
L.~P.~Kouwenhoven, 
C.~M.~Marcus, P.~L.~McEuen, S.~Tarucha, R.~M.~Westervelt, 
and N.~S.~Wingreen, 
in {\it Mesoscopic Electron Transport},  
edited by L.~L.~Sohn, L.~P.~Kouwenhoven, and G.~Sch\"on 
(Kluwer, Dordrecht, 1997) pp. 105-214. 
 
\bibitem{Stopa98} 
M.~Stopa, 
{Physica B} {\bf 251}, 228 (1998). 
 
\bibitem{Chang96} 
A.~M.~Chang,  
H.~U.~Baranger, L.~N.~Pfeiffer, K.~W.~West, and T.~Y.~Chang, 
{ Phys. Rev. Lett.} {\bf 76}, 1695 (1996). 
 
\bibitem{MarcusFolk96} 
 J.~A.~Folk, S.~R.~Patel, S.~F.~Godijn, A.~G.~Huibers,  
S.~M.~Cronenwett, and C.~M.~Marcus, 
{ Phys. Rev. Lett.} {\bf 76}, 1699 (1996). 
 
\bibitem{Alhassid98}  
Y.~Alhassid, M.~G\"ok\c{c}edag, and A.~D.~Stone, 
{ Phys. Rev. B} {\bf 58}, 7524 (1998). 
 
\bibitem{MarcusPatel98} 
S.~R.~Patel, 
D.~R.~Stewart, C.~M.~Marcus, M.~G\"ok\c{c}edag, Y.~Alhassid, A.~D.~Stone, 
C.~I.~Duru\"oz, and J.~S.~Harris,~Jr., 
{ Phys. Rev Lett.} {\bf 81}, 5900 (1998). 
 
\bibitem{Stopa96} 
 M.~Stopa, 
{ Phys. Rev. B} {\bf 54}, 13767 (1996). 
 
\bibitem{Hackenbroich97} 
G.~Hackenbroich, W.~D.~Heiss, and H.~A.~Weidenm\"uller, 
{ Phys.~Rev.~Lett.} {\bf 79}, 127 (1997). 
 
\bibitem{Vallejos} 
R.~O.~Vallejos, C.~H.~Lewenkopf, and E.~R.~Mucciolo,
Phys.~Rev.~B {\bf 60}, 13682 (1999). 
  
\bibitem{Gutzwiller} 
M.~Gutzwiller, {\it Chaos in Classical and Quantum Mechanics} 
(Springer, New York, 1990). 
 
\bibitem{BeimsDelos98}  
J.~B.~Delos and C.~D.~Schwieters, 
in {\it Classical, Semiclassical and Quantum Dynamics in Atoms}, 
 ed. by 
 H.~Friedrich and B.~Eckhardt, (Springer, Berlin, 1997) p. 223-247; 
M. W. Beims, V. Kondratovich, and J. B. Delos, 
Phys. Rev. Lett. {\bf 81}, 4537 (1998). 
 
\bibitem{heller} E.~J.~Heller, in  
{\it Chaos and Quantum Physics}, edited by M.~J.~Giannoni, 
 A.~Voros, and J.~Zinn-Justin (Elsevier, Amsterdam, 1991) pp. 547-663. 
 
\bibitem{KaplanHeller}  
L.~Kaplan, 
{ Phys. Rev. Lett} {\bf 80}, 2582 (1998); 
L.~Kaplan, and E.~J.~Heller,  
{ Annals of Phys.} {\bf 264}, 171 (1998). 
 
\bibitem{antidots} 
D.~Weiss, K.~Richter, A.~Menschig, R.~Bergmann, H.~Schweizer,  
K.~von~Klitzing and G.~Weimann,  
Phys. Rev. Lett. {\bf 70}, 4118 (1993); 
G.~Hackenbroich, and F.~von~Oppen, Europhys. Lett., {\bf 29}, 151 (1995); 
K.~Richter, Europhys. Lett. {\bf 29}, 7 (1995). 
 
\bibitem{e2n_prl} 
E.~E.~Narimanov, A.~D.~Stone, and G.~S.~Boebinger, 
{ Phys. Rev. Lett.} {\bf 80}, 4024 (1998); an 
alternative semiclassical theory of resonant magnetotunneling 
was developed by E.~B.~Bogomolny and  
D.~C.~Rouben, { Europhys. Lett.} {\bf 43}, 
111 (1998). 
 
\bibitem{monteiro} 
T.~S.~Monteiro, D.~Delande, A.~J.~Fisher, G.~S.~Boebinger, 
{ Phys. Rev. B} {\bf 56}, 3913 (1997). 
 
\bibitem{saraga98}  
D.~Saraga and T.~S.~Monteiro, 
Phys. Rev. Lett. {\bf 81}, 5796 (1998). 
 
\bibitem{e2n_physicad} E.~E.~Narimanov, and A.~D.~Stone,  
Physica D {\bf 131}, 220 (1999). 
 
\bibitem{MarcusCronenwett97} 
S.~M.~Cronenwett, S.~R.~Patel, C.~M.~Marcus, K.~Campman, and A.~G.~Gossard, 
{ Phys. Rev. Lett.} {\bf 79}, 2312 (1997). 
 
\bibitem{Nar_cbdyn1} 
E.~E.~Narimanov, N.~R.~Cerruti, H.~U.~Baranger, and S.~Tomsovic, 
Phys. Rev. Lett. {\bf 83}, 2640 (1999). 

\bibitem{beenakker} 
 C.~W.~J.~Beenakker,  
{ Phys. Rev. B} {\bf 44}, 1646 (1991). 
 
\bibitem{bardeen} 
J.~Bardeen, 
{ Phys. Rev. Lett.} {\bf 6}, 57 (1961). 
 
\bibitem{Berry77} M.~V.~Berry, J.~Phys.~A {\bf 10}, 2083 (1977);
for a description of the properties of random functions see also
P.~W.~O'Connor, J.~Gehlen, and E.~J.~Heller, 
Phys.~Rev.~Lett. {\bf 58}, 1296 (1987). 
 
\bibitem{Alhassid95} Y.~Alhassid and C.~H.~Lewenkopf,  
Phys.~Rev.~Lett. {\bf 75}, 3922 (1995). 
 
\bibitem{Srednicki96} M.~Srednicki, Phys. Rev. E {\bf 54}, 954 (1996). 
 
\bibitem{Prigodin93} V.~N.~Prigodin, K.~B.~Efetov, and S.~Idia, 
Phys.~Rev.~Lett. {\bf 71}, 1230 (1993). 
 
\bibitem{Prigodin94} V.~N.~Prigodin, B.~L.~Altshuler, K.~B.~Efetov,  
and S.~Idia, Phys.~Rev.~Lett. {\bf 72}, 546 (1994). 
 
\bibitem{Prigodin95} V.~N.~Prigodin, 
Phys.~Rev.~Lett. {\bf 74}, 1566 (1995). 
 
\bibitem{PrigodinSridhar95} V.~N.~Prigodin, N.~Taniguchi, A.~Kudrolli, 
V.~Kidambi, and S.~Sridhar, Phys.~Rev.~Lett. {\bf 75}, 2392 (1995). 
 
\bibitem{SrednickiStiernelof96} M.~Srednicki and F.~Stiernelof,  
J.~Phys.~A {\bf 29}, 5817 (1996). 
 
\bibitem{Hortikar98} S.~Hortikar and M.~Srednicki, 
Phys.~Rev.~Lett. {\bf 80}, 1646 (1998). 
 
\bibitem{Kaplan} 
L.~Kaplan, Phys. Rev. E {\bf 62}, 3476 (2000) [arXiv:nlin.CD/0003013 (2000)]. 
 
\bibitem{maxent} 
C.~Jarzynski, Phys. Rev. E {\bf 56}, 2254 (1997).

\bibitem{Shannon48} C.~E.~Shannon, {\it A Mathematical Theory 
of Communication}, The Bell System Technical Journal {\bf  
27}, 379-423, 623-656 (1948). 
 
\bibitem{Reichl} L.~E.~Reichl, {\it The Transition to Chaos In 
Conservative Classical Systems: Quantum Manifestations} 
(Springer, New York, 1992). 
 
\bibitem{BroOreHal} 
P. W. Brouwer, Y. Oreg, and B. I. Halperin, 
Phys. Rev. B {\bf 60}, R13977 (1999). 
 
\bibitem{BarUllGla} 
H. U. Baranger, D. Ullmo, and L. I. Glazman, 
Phys. Rev. B {\bf 61}, R2425 (2000). 
 
\bibitem{UllBar_inprep} D.~Ullmo and H.~U.~Baranger, in preparation. 
 
\bibitem{Srednicki_pr_com} M.~Srednicki, private communication. 
 
\bibitem{Julie} 
T.~Szeredi, J.~H.~Lefebvre, D.~A.~Goodings, 
{ Nonlinearity} {\bf 7}, 1463 (1994). 
 
\bibitem{Brack} M.~Brack, O.~Genzken, and K.~Hansen,  
Z. Phys. D---Atoms, Molecules and Clusters {\bf 21}, 65 (1991). 
 
\bibitem{Marcus_comm} C.~M.~Marcus, private communication. 

\bibitem{correction} This estimate and conclusion contradicts that in
our previous publication Ref. \protect\onlinecite{Nar_cbdyn1} which
we now believe to be missing a factor of $\pi$.

\bibitem{Tomsovic00}
M.~A.~Topinka, B.~J.~LeRoy, R.~M.~Westervelt, S.~E.~J.~Shaw, R.~Fleischmann, 
E.~J.~Heller, K.~D.~Maranowski and A.~C.~Gossard, arXiv:cond-mat/0010348; 
see also M.~A.~Wolfson and S.~Tomsovic, J. Acoust. Soc. Am. {\bf 105}, 
1116 (1999); J. Acoust. Soc. Am. in press [arXiv:nlin.CD/0002030]. 
 
\bibitem{StewartMarcus} D.~R.~Stewart,  
D.~Sprinzak, C.~M.~Marcus, C.~I.~Duru\"oz, and J.~S.~Harris,  
Science {\bf 278}, 1784 (1997). 
 
\end{references}
\end{document}